\documentclass[twocolumn,showpacs,preprintnumbers,amsmath,amssymb]{revtex4}
\usepackage{tabularx,graphicx}\begin{document}
%\documentstyle[aps]{revtex}
%\documentstyle[preprint,aps]{revtex}
%\begin{document}
\newcommand{\beq}{\begin{equation}}
\newcommand{\eeq}{\end{equation}}
\newcommand{\beqn}{\begin{eqnarray}}
\newcommand{\eeqn}{\end{eqnarray}}
\newcommand{\bmath}{\begin{subequations}}
\newcommand{\emath}{\end{subequations}}

%\newcommand{\bmath}{\begin{mathletters}}
%\newcommand{\emath}{\end{mathletters}}
%\draft
%\twocolumn[\hsize\textwidth\columnwidth\hsize\csname @twocolumnfalse\endcsname 
\title{ The fundamental role of charge asymmetry in superconductivity}
\author{J.E. Hirsch}
\address{Department of Physics, University of California, San Diego\\
La Jolla, CA 92093-0319}

\date{\today} 

\begin{abstract}
Neither BCS theory nor London theory contain any charge asymmetry. However it is an experimental fact that a rotating superconductor always exhibits a magnetic field parallel, never antiparallel, to its angular velocity. This and several other experimental observations point to a special  role of charge asymmetry in superconductivity, which is the foundation of the theory of hole superconductivity. The theory  describes heavy dressed {\it  positive}  hole carriers in the normal state that undress by pairing and become light undressed {\it  negative} electron carriers in the superconducting state. Superconductivity is driven by kinetic energy lowering rather than by electron-phonon coupling as in BCS.  In quantum mechanics, kinetic energy lowering is associated with $expansion$ of the electronic wave function, and   hence we predict:  (1) Superconductors expel $negative$ charge from their interior which consequently becomes $positively$ charged; (2) Macroscopic electrostatic fields exist in the interior of superconductors always, and in certain cases also outside near the surface; (3) Macroscopic  spin currents exist in the superconducting state; (4) Superconductors are 'rigid' with respect to their response to applied longitudinal electric fields.  These predictions apply to all superconductors and are testable but are as yet untested.  The theory predicts highest $T_c$'s for materials for which normal state transport occurs through $(positive)$ holes in $negatively$ charged anions.

 \end{abstract}
\maketitle
\section{Charge asymmetry manifestations}
The fundamental charge asymmetry of matter manifests itself in the fact that the
negative electron is 2000 times lighter than the positive proton. However normal metals do not clearly
display this charge asymmetry,
since their transport properties can be sometimes understood as originating from mobile $negative$ carriers (electrons) and
sometimes from mobile $positive$ carriers (holes).

The observation  that superconductivity   occurs predominantly in materials where
 the carriers in the normal state are holes
rather than electrons was made  long ago\cite{chap}. It has been further reinforced by the finding that both in
high $T_c$ cuprates and in $MgB_2$ the carriers that drive superconductivity appear to be hole-like (this also appears to be the case in the electron-doped cuprates\cite{electron}). Other manifestations of charge asymmetry in superconductors are the sign reversal of the Hall coefficient right below $T_c$ (it goes from positive in the normal state to negative in the superconducting
state)\cite{hall} and the voltage asymmetry in NIS tunneling in high $T_c$ cuprates (higher conductance for negatively
biased sample)\cite{tunn,andersonong}.

Furthermore it has been know theoretically since 1933\cite{becker} and experimentally since  1964\cite{hildebrand}  that  rotating superconductors posess  a magnetic field in their interior that gives rise to a magnetic moment that points always {\it in the same direction} as their angular velocity, as depicted in Fig. 1.  Qualitatively this can be understood as resulting from the fact that the 
{\it negative} superfluid lags behind when the {\it positive} ions are rotating, and hence indicates that the superfluid
carriers are always {\it negatively} charged. The importance of this observation has not yet been widely recognized\cite{eha}.

\begin{figure}
\resizebox{6cm}{!}{\includegraphics[width=9cm]{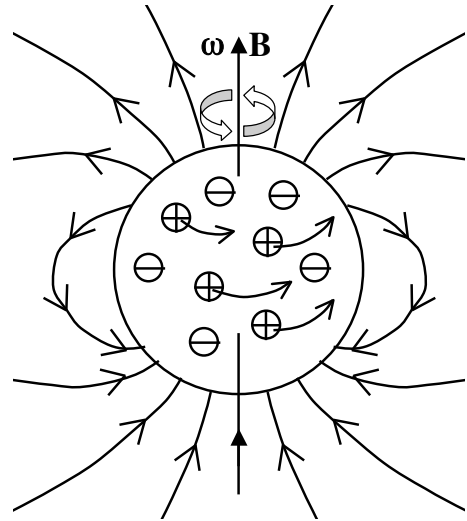}}
%\resizebox{8cm}{!}{\includegraphics[width=9cm]{fig2.pdf}}
\caption{Experimental proof that superconductors know about charge asymmetry. The positive charge in a rotating superconductor
rotates faster than the negative charge, giving rise to a magnetic moment parallel   to the angular velocity.}
\label{fig2}
\end{figure}

The theory discussed here implies, in agreement with these observations, that in the transition to the superconducting state 
{\it hole-like normal state carriers} become 
{\it electron-like superconducting carriers}.  Our theory\cite{hole} has in common with BCS theory that superconductivity 
occurs through pairing of time-reversed carriers, and with London theory that the Meissner effect originates in
macroscopic quantum coherence.   There are however significant differences with the conventional theories.

\section{Holes and antibonding electrons}

If  the concept of holes had never been invented\cite{heisenberg}, understanding of superconductivity would perhaps have come much
easier. In a metal where transport is said to occur through holes, the
carriers at the Fermi energy are in fact antibonding electrons. It is the antibonding electrons at the Fermi energy that are the
key to superconductivity in our theory.

Electrons in energy bands have increasingly higher energy as more electrons are added to the band, due to the
Pauli exclusion principle. Bose-condensed particles attain the lowest possible energy state. It is only natural
to conclude that superconductivity becomes more favorable when the electrons at the Fermi level are farthest away
from the bottom of the band, because it is then that electrons at the Fermi energy gain the most by Bose-condensing.
Because only bosons can Bose-condense, antibonding electrons have to go from half-integer spin
to integer-spin particles, hence they have to pair.

\begin{figure}
%\resizebox{8.5cm}{!}{\includegraphics[width=7cm]{fig4.pdf}}
\resizebox{8.5cm}{!}{\includegraphics[width=7cm]{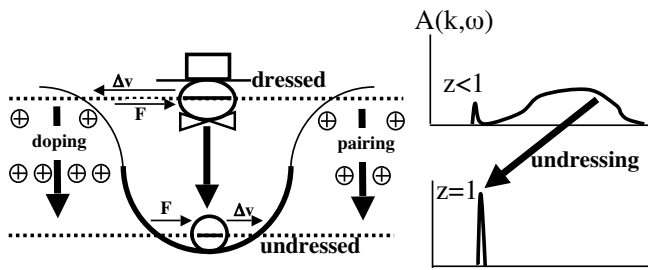}}
\caption{When the Fermi level is close to the bottom of the band, carriers at the Fermi energy are undressed electrons. Their spectral weight  is all 
in the quasiparticle $\delta-$function, and when an external force $F$ is applied they respond with a change in velocity $\Delta v$ in the same 
direction as the applied force. Instead when the Fermi level is close to  the top of the band quasiparticles at the Fermi energy are dressed: they respond with a change in 
velocity opposite to the applied external force, and their spectral weight is mostly in the high frequency incoherent part of the
spectral function $A(k,\omega)$. Upon pairing these quasiparticles undress and spectral weight is transfered from high to low frequencies. Doping of holes or pairing of holes lead to an increase in the hole concentration around a given hole and
to undressing.}
\label{atom4}
\end{figure}

When the Fermi level is close to the top of the band, carriers at the Fermi energy (antibonding electrons) are 'dressed' by the electron-ion
interaction and as a consequence have a $negative$ effective mass. This means that when an external force acts on
those electrons (eg through an applied electric field) they acquire momentum in direction opposite to the force, and the ionic
lattice picks up the difference. As a consequence those electrons oppose, rather than contribute to, the electrical conductivity 
of the metal.

Antibonding electrons are also 'dressed' by the electron-electron interaction. This causes the magnitude of the effective mass to be large
and the quasiparticle weight to be small and further inhibits the electrical conductivity. Dynamic Hubbard models\cite{dynhub,dynhub2} and electron-hole asymmetric polaron models\cite{polaron} describe
the increased dressing of carriers by the electron-electron interaction as the Fermi level rises from the bottom to the top of a band.
These model Hamiltonians have a simple and direct connection with the physics of real electrons in atomic orbitals of varying occupation\cite{holeelec}.
They predict that for favorable model parameters, in particular corresponding to the case when the ions are anion-like, superconductivity
occurs driven by 'undressing' of carriers when they pair:  despite the cost in Coulomb energy a larger reduction of kinetic energy occurs,
associated also with an increase in the quasiparticle weight, giving rise to a positive condensation energy.

Summarizing, the dressing of the quasiparticles at the Fermi energy from both electron-ion and electron-electron interactions increases as
the Fermi level rises in the band, and inhibits the electrical conductivity in the normal state. Pairing of holes increases the local hole concentration, thus mimicking a situation where the band is less full, and  hence should lead to  'undressing' from both the electron-electron and the electron-ion interaction, as shown schematically in Fig. 2,  and to uninhibited electrical conductivity.

\section{Superconductivity from 'undressing'}

It is indeed seen experimentally that superconductivity is favored ($T_c$ is highest) when the carriers in the normal state are
hole-like and heavily dressed and the electrical conductivity is small, and that 'undressing' occurs upon transition to the superconducting state. Photoemission experiments in high $T_c$ cuprates
show a sharp increase in the quasiparticle weight as the system goes superconducting\cite{ding}, optical experiments show lowering of kinetic
energy as the system goes superconducting\cite{optical}, and the sign of the Hall coefficient changes from positive to negative as the system goes superconducting\cite{hall}. Experiments also show that 'undressing'
  occurs upon doping of holes  in the normal  state: the quasiparticle weight increases\cite{normal}, 
  optical spectral weight is transfered from high to low frequencies\cite{uchida}, and the sign of the Hall coefficient
  changes from positive to negative\cite{takagi}. When the hole concentration becomes   large carriers  in the normal state are already undressed so
  undressing through pairing can no longer occur and superconductivity dissappears.  The situation is schematically depicted in Fig. 3.
  
  \begin{figure}
\resizebox{7cm}{!}{\includegraphics[width=9cm]{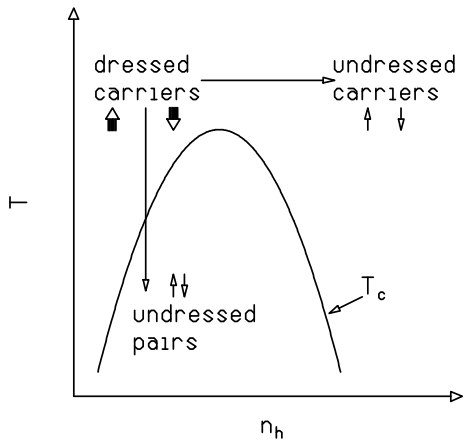}}
%\resizebox{8cm}{!}{\includegraphics[width=9cm]{fig2.pdf}}
\caption{$T_c$ versus hole concentration $n_h$. For $n_h=0$ the band is full. When the local hole concentration around a hole increases
either through doping or through pairing, holes undress. From its maximum, $T_c$ decreases
to the left as the number of carriers goes to zero and to the right as the carriers become  undressed in the
normal state and no longer benefit from pairing. }
\label{fig2}
\end{figure}

 Dynamic Hubbard models naturally describe this 'undressing'. Using a Lang-Firsov transformation on the microscopic Hamiltonians, the relation between bare (original) operators $c_{i\sigma}^\dagger$
 and transformed (quasiparticle) operators $\tilde{c}_{i\sigma}^\dagger$ is, in a hole representation\cite{undressing}  
 \beq
 c_{i\sigma}^\dagger=  \frac{1+\Upsilon \tilde{n}_{i,-\sigma}}{1+\Upsilon} \tilde{c}_{i\sigma}^\dagger
 \eeq
 with  $\Upsilon>0$ and $\tilde{n}_{i\sigma}=\tilde{c}_{i\sigma}^\dagger\tilde{c}_{i\sigma}$. 
 The 'undressing parameter' Upsilon ($\Upsilon>0$) 
 describes the undressing of holes with increasing local hole concentration and drives the transition to superconductivity through the enhancement of the hopping amplitude $\Delta t=\Upsilon t_h$, with $t_h$ the single hole hopping amplitude\cite{holesc},
 as well as the undressing upon hole doping in the normal state. 
 High (low) $T_c$ materials are described by large (small) values of $\Upsilon$, and the
 dressing of quasiparticles in the normal state is an increasing function of $\Upsilon$.

However nature is bolder than what the model Hamiltonians proposed so far suggest.  Experiments on rotating superconductors show that the magnetic field in their interior generated for
rotation frequency $\vec{\omega}$ is\cite{becker,hildebrand}
\beq
\vec{B}=-\frac{2m_ec}{e}\vec{\omega}
\eeq
with $m_e$ and $e$ the $bare$ electron mass and charge. This indicates that superfluid electrons in superconductors, just as in atoms, behave as 
  totally bare undressed electrons (except for the pairing correlations). Hence the dressed quasiparticles (holes) in the normal state become bare undressed electrons
  in the superconducting state\cite{eha}. 

\section{Negative charge expulsion}

In cuprates the NIS tunneling current is observed to be larger for a negatively biased sample, i.e. when electrons tunnel
$out$ of the superconductor. The theory of hole superconductivity predicts that this is so for all superconductors
due to the finite slope of the BCS gap function $\Delta_k=\Delta(\epsilon_k)$, with $\epsilon_k$ the hole
band energy.\cite{tunn} The gap function versus $\epsilon_k$ and the quasiparticle energy
$E_k=\sqrt{(\epsilon_k-\mu)^2+\Delta_k^2}$ are depicted in Fig. 4. The asymmetry in the magnitude of the tunneling
current suggests that superconductors have a tendency to lose electrons and  become positively  charged.

\begin{figure}
\resizebox{7cm}{!}{\includegraphics[width=9cm]{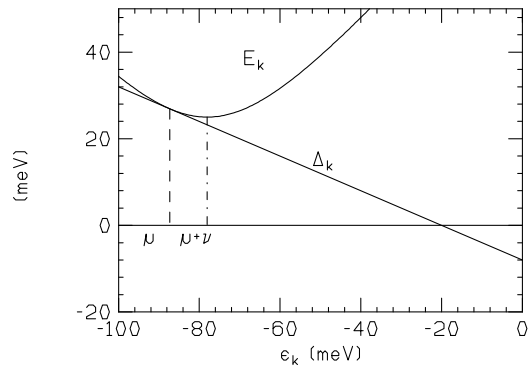}}
%\resizebox{8cm}{!}{\includegraphics[width=9cm]{fig2.pdf}}
\caption{Gap fuction $\Delta_k$ and quasiparticle energy $E_k$ versus
hole band energy $\epsilon_k$ in the model of hole superconductivity. The minimum quasiparticle energy occurs at $\epsilon_k=\mu+\nu$. $\mu$ is the chemical potential
and $\nu$ is proportional to the gap function slope. }
\label{fig2}
\end{figure}

The minimum quasiparticle excitation energy occurs at the quasiparticle chemical potential $\mu'=\mu+\nu$ rather than at the 
chemical potential of the condensate $\mu$. As a consequence superfluid electrons are expelled from the interior of the 
superconductor, tending to equalize the two chemical potentials.\cite{chargeexp} This can also be seen 
through the temperature dependence of
quasiparticle weights for electron and hole creation below $T_c$\cite{undressing}. The expelled negative charge 
accumulates near the surface and the balance between Coulomb charging energy cost and condensation energy gain 
determines the amount of negative charge expelled\cite{charge}. For samples of dimension much larger than the
penetration depth, the density of negative charge near the surface $\rho_-$ is independent of sample size, while the concentration of positive charge in the interior, $\rho_0$, decreases with increasing sample size. A qualitative picture
of the charge distribution in a spherical superconductor is shown in Fig. 5. As in metal clusters one may expect some 
negative charge to spill out beyond the surface of the superconductor.\cite{atom}

\begin{figure}
\resizebox{6cm}{!}{\includegraphics[width=9cm]{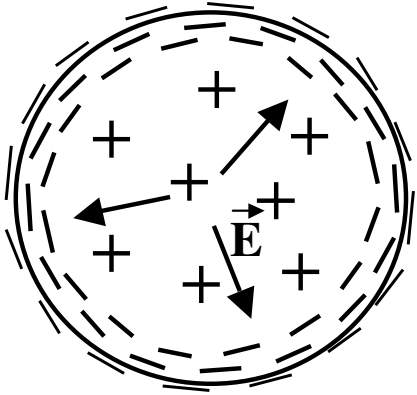}}
%\resizebox{8cm}{!}{\includegraphics[width=9cm]{fig2.pdf}}
\caption{Schematic picture of a spherical superconducting body. Negative charge is expelled from the bulk to the surface
and an outward-pointing electric field exists in the interior. Some negative charge spills out beyond the surface of
the superconductor. }
\label{fig2}
\end{figure}

\section{Macroscopic electrodynamics}

The predictions of the microscopic theory have a remarkably simple macroscopic electrodynamic description\cite{electrodyn}. The existence of the pairing gap implies the  validity of the conventional London equation
\bmath
\beq
\vec{J}=-\frac{c}{4\pi\lambda_L^2}\vec{A}
\eeq
with $\vec{A}$ the magnetic vector potential and $\lambda_L$ the London penetration depth.
  However unlike the conventional theory we assume that $\vec{A}$ in Eq. (3a) obeys the
Lorenz gauge
\beq
\vec{\nabla}\cdot\vec{A}=-\frac{1}{c}\frac{\partial \phi}{\partial t}
\eeq
\emath 
with $\phi$ the scalar potential, instead of the conventional London gauge. Eqs (3) give rise to a new description of the
electrodynamics of superconductors\cite{electrodyn}.  The charge density and electrostatic potential are related by 
\beq
\rho -\rho_0=-\frac{1}{4\pi\lambda_L^2}(\phi-\phi_0)
\eeq 
and the electrostatic equations for the charge density and electric field are
\bmath
\beq
\nabla^2 (\rho- \rho_0)=\frac{1}{\lambda_L^2}(\rho-\rho_0) 
\eeq
\beq
\nabla^2(\vec{E}-\vec{E}_0)=\frac{1}{\lambda_L^2}(\vec{E}-\vec{E_0})
\eeq
\emath
with $\phi_0$ the electrostatic potential resulting from a uniform positive charge density $\rho_0$ throughout the volume of
the superconductor, and $\vec{E}_0=-\vec{\nabla}\phi_0$.  Eq. (5) implies that the expelled negative charge resides within a London penetration depth of the
surface of the superconductor.

The electrodynamics of a superconductor is described by the remarkably simple four-dimensional covariant equation
\beq
\Box^2({\it{A}}-{\it{A_0)}}=\frac{1}{\lambda_L^2}({\it{A}}-{\it{A_0}})
\eeq
with
\bmath
\beq
\Box^2=\nabla^2-\frac{1}{c^2}\frac{\partial^2}{\partial t^2}
\eeq
\beq
{\it{A}}=(\vec{A}(\vec{r},t),i\phi(\vec{r},t))
\eeq
\beq
{\it{A_0}}=(0,i\phi_0(\vec{r}))
\eeq
\emath
The form Eq. (7c) is only valid in the rest frame of the superconductor. Eq. (6) can be 'derived' by assuming rigidity
of the wavefunction of the superfluid in Klein Gordon theory\cite{electrodyn}.
The frequency dependent dielectric constant describing the superfluid response has the remarkably simple form
\beq
\epsilon_s(q,\omega)=\frac{\omega_p^2+c^2q^2-\omega^2}{c^2q^2-\omega^2}
\eeq
with $\omega_p=c/\lambda_{L}$. 

There are several experimentally testable consequences of these equations. Applied longitudinal electric fields should
penetrate inside superconductors a distance $\lambda_L$ rather than the much shorter Thomas-Fermi length. To test this
effect the temperature should be low enough that very few quasiparticles are excited. We estimate that for
temperatures lower than $T_c/20$ a measurable change (increase) in the capacitance of a capacitor with superconducting 
plates should be seen upon application of a magnetic field large enough to destroy the superconductivity.

\begin{figure}
\resizebox{6cm}{!}{\includegraphics[width=9cm]{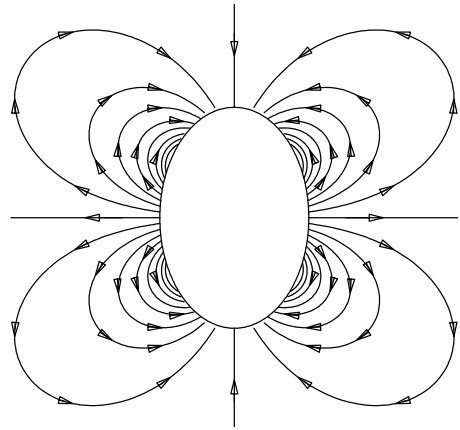}}
%\resizebox{8cm}{!}{\includegraphics[width=9cm]{fig2.pdf}}
\caption{Electric field lines for an ellipsoidal superconductor with ratio of semimajor to semiminor axis of $1.5$}
\label{fig2}
\end{figure}

The charge distribution and electrostatic field can be calculated analytically for a spherical geometry and numerically in
other cases. When the superconductor is not spherical an electric field exists $outside$ the superconductor.
Fig. 6 shows the electric field lines for a superconductor of ellipsoidal shape. This should be experimentally
detectable in small samples\cite{ellipsoid}.

The plasmon dispersion relation that results from Eq. (8)
\beq
\omega_s^2(q)=\omega_p^2+c^2q^2
\eeq
differs from the plasmon dispersion relation in the normal state
\beq
\omega_n^2(q)=\omega_p^2+\frac{3}{5}v_F^2q^2
\eeq
in that it predicts a much steeper dispersion for the superfluid ($v_F=$Fermi velocity). This should be
seen in electron energy loss spectroscopy. Related to this, surface plasmons in small metallic particles should show a 
blue shift in the superconducting state at low temperatures\cite{electrodyn}.

\section{Spin currents}

Another remarkable prediction of our theory is the existence of macroscopic spin currents in the  superconducting state in the
absence of applied electric and magnetic fields\cite{spinc}. In the early days of superconductivity superconductors were
expected to have macroscopic charge currents in their ground state\cite{heis}. This was proven impossible
by a 'Bloch theorem'\cite{bohm}, however that theorem does not apply to spin currents.

It is interesting to note that a Cooper pair $c_{k\uparrow}^\dagger c_{-k\downarrow}^\dagger$ carries a spin current and not
a charge current. A spin current will result if Cooper pairs  $(k\uparrow,-k\downarrow)$ and $(-k\uparrow,k\downarrow)$ have different amplitudes. 
Anderson\cite{anderson} pointed out that superconductors should be insensitive to scattering by 
nonmagnetic impurities because electrons in a Cooper pair are in time reversed states. By the same argument, 
because a spin current does not break time reversal invariance it will not be degraded by potential scattering 
and will persist as long as the system is in the superconducting state. There is no way to stop it by any
external perturbation. 

Because a macroscopic electric field is predicted to exist in the interior of superconductors in our theory, the existence of
macroscopic spin currents necessarily follows. The spin-orbit interaction energy
\beq
U_{so}=\frac{e}{2m_ec^2}\vec{S}\cdot (\vec{v}\times \vec{E})
\eeq
will be lowered by electrons orbiting predominantly such that their orbital angular momentum is parallel to their magnetic moment,
as shown schematically in Fig. 7.
\begin{figure}
\resizebox{6cm}{!}{\includegraphics[width=9cm]{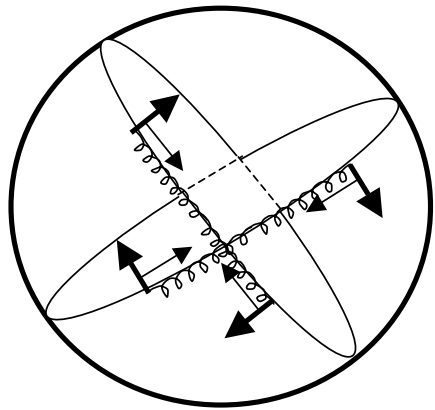}}
%\resizebox{8cm}{!}{\includegraphics[width=9cm]{fig2.pdf}}
\caption{Cooper pairs in the presence of a macroscopic electric field pointing outward will orbit predominantly so that their orbital 
angular momentum is parallel to the electron magnetic moment giving rise to a macroscopic spin current.}
\label{fig2}
\end{figure}
Remarkably, the microscopic theory  also shows that  the superconducting
condensation energy increases in the presence of spin-orbit splitting when the pairing is driven by lowering of kinetic energy as in the
model of hole superconductivity\cite{spinc}.

Experiments that may be able to verify the existence of macroscopic spin currents in superconductors are spin-polarized
neutron scattering\cite{neutrons} and angle resolved photoemission using circularly polarized light\cite{varma}. Furthermore 
we expect that in the presence of an applied
 magnetic field, spin currents will induce a quadrupolar electric field around a superconductor
that should be experimentally detectable.

\section{Superconductors as giant atoms}

It is well established that superconductors exhibit quantum coherence at a macroscopic scale. That is, the phase of the
wavefunction is well defined over the entire volume of the superconductor. As first envisaged by London, the London
equation can be understood by regarding a superconductor as a 'giant atom'\cite{london}. The theory discussed here
carries the analogy one step further: not only is a superconductor a giant atom as far as its diamagnetic response is
concerned, but just as an atom it has more positive charge near its
center and more negative charge near its boundaries. Pairing of time-reversed electrons and an energy gap are still
part of the theory as in the conventional one, but charge asymmetry is its most fundamental ingredient. 
The various indications of charge asymmetry in superconductivity seen
so far in experiment such as the asymmetry in NIS tunneling  
are dwarfed by a much more dramatic manifestation of the fundamental charge asymmetry of matter:
 the macroscopically inhomogeneous charge distribution of positive and negative charge,
 qualitatively mirroring the one at the atomic level, that we predict to exist
 in all superconductors.
  
 Local charge neutrality in normal metals results from
minimization of the mobile electron {\it potential} energy. However if quantum mechanics acts on a macroscopic scale in superconductors
 the
wavefunction is determined by minimization of the potential $plus$ $kinetic$ energy. When the normal metal goes superconducting, 
expansion of the electronic wave function leads to kinetic energy lowering only partially compensated by an increase in 
potential energy, resulting in the inhomogeneous charge distribution depicted in Fig. 5.

The existence of a macroscopic electric field inside superconductors can be deduced from Eq. (2) without necessity to invoke
the microscopic theory. According to Larmor's theorem\cite{larmor}, electrons in the rest frame of a rotating superconductor
  experience a Coriolis force
that is compensated by the magnetic field Eq. (2)\cite{alben}. However, {\it Larmor's theorem is only valid when the rotation frequency
$\omega$ is much $smaller$ than the intrinsic frequency of motion of the electron}\cite{larmor}. Thus one is led to conclude that 
electrons in the superconductor  traverse macroscopic orbits with   angular velocity much larger than any rotation frequency of
the superconducting body in a laboratory experiment, from which the existence of an internal macroscopic electric field follows.

However, it is possible that even after experiments determine that the charge distribution in superconductors is
as predicted here, it will still be mantained that pairing   is driven by electron-phonon interaction rather than by undressing. 
Such a statement cannot be
proven wrong by experiment because it is non-falsifiable, so one can only hope that it will become increasingly
irrelevant and in time fade away. The theory discussed here predicts that favorable material conditions  to superconductivity are
{\it hole conduction through negatively charged anions} and that any correlation with electron-phonon parameters is
a $consequence$ of those conditions.

The theory discussed here has many testable predictions, as discussed here and elsewhere. In particular the
electrostatic field distribution around superconductors can be calculated for any given shape of the
superconducting body. The field line configuration is predicted with no adjustable parameters, thus providing
a stringent test of the theory. In particular, as seen in the example of Fig. 6, electric field lines always go $in$ in regions of
high curvature and go $out$ in regions of small curvature.

The theory discussed here applies to all superconductors and provides answers to many puzzles. The Meissner effect
can be understood to arise from the Lorentz force on the expelled radially outgoing electrons when the system is cooled
from above to below $T_c$ in the presence of a magnetic field\cite{lorentz}. Similarly the generation of spin currents can be understood
from a spin Hall effect on the expelled electrons\cite{spinhall1,spinhall2,spinhall3,spinc}. 
Similarly the generation of the London field Eq. (2) when a rotating
normal metal is cooled below $T_c$, regarded as
'quite absurd' by London\cite{absurd} , can be understood from the Coriolis force acting on the expelled electrons\cite{lorentz}.
The essential physics of superconductivity is predicted to be the same for high $T_c$ cuprates, $MgB_2$, heavy fermion superconductors,
organic superconductors, conventional superconductors, as well as for all other superconducting materials.

\end{document}